\documentclass[epj,english,twocolumn]{webofc}
\usepackage[varg]{txfonts}   
\usepackage{hyperref}        
\usepackage{soul}
\usepackage{wasysym}
\woctitle{Heavy Ion Accelerator Symposium 2019}
\begin{document}
\title{Neutron stars from crust to core within the Quark-meson coupling model}
%
%

\author{S.Anti\'{c}\inst{1}\fnsep\thanks{\email{sofija.antic@adelaide.edu.au}} \and J. R. Stone\inst{2,3} \and A. W. Thomas\inst{1}}

\institute{CSSM, Department of Physics, University of Adelaide, SA 5005 Australia
\and
           Department of Physics (Astro), University of Oxford, OX1	3RH United Kingdom
\and
           Department of Physics and Astronomy, University of Tennessee, TN 37996 USA
          }

\abstract{%
Recent years continue to be an exciting time for the neutron star physics, providing many new observations and insights to these natural ’laboratories’ of cold dense matter. To describe them, there are many models on the market but still none that would reproduce all observed and experimental data. The quark-meson coupling model stands out with its natural inclusion of hyperons as dense matter building blocks, and fewer parameters necessary to obtain the nuclear matter equation of state. The latest advances of the QMC model and its application to the neutron star physics will be presented, within which we build the neutron star's outer crust from finite nuclei up to the neutron drip line. The appearance of different elements and their position in the crust of a neutron star is explored and compared to the predictions of various models, giving the same quality of the results for the QMC model as for the models when the nucleon structure is not taken into account.
}
\maketitle
\section{Introduction} \label{sec:1}

The neutron stars (NS) are in general very complex objects that connect many different fields of research, from nuclear and particle physics and astrophysics to general relativity. Small in size, with the radii less than 14km and masses as high as two solar mass, these objects are among the densest in the Universe. Starting from their outer layer, the crust, density of stellar matter increases very rapidly as we travel inwards to the deeper layers of a NS due to the very strong gravitational force which is around $10^{11}$ times stronger than what we experience on Earth. For that reason the dilute atmosphere with $\rho < 10^{-15}$ fm$^{-3}$ ($\rho < 10$ g/cm$^3$) is only around 100 mm thick while the full crust below, with densities in the range $10^{-15}$ fm$^{-3} < \rho < 10^{-2}$ fm$^{-3}$, accounts for around $1\%$ of the total NS radius. 

The composition of  NS's outer crust starts with iron nuclei arranged in a body-centred cubic lattice emerged in an electron free gas. With the increase of density and the distance from the surface, nuclei in the lattice become more neutron rich. This continues up to a point where nuclei do not have any free energy levels left to accommodate additional neutrons, leading to the appearance of neutron gas. The nuclear lattice is now immersed in the already present free electron gas and additionally neutron gas, who's presence marks the transition to the inner crust of a NS at density $\rho \sim 10^{-5}$ fm$^{-3}$. The inner crust has more complicated structure, since nuclei eventually start to cluster and form exotic shapes, commonly addressed as nuclear pasta. At densities $\rho > 10^{-2}$ fm$^{-3}$ the transition into homogeneous nuclear matter is happening, composed of neutrons, protons, electrons and muons, indicating the beginning of what we call the NS core. For an average NS with $M \sim 1.4 M_{\astrosun}$ the nearly-pure neutron matter with admixture of protons and electrons is all there is in a NS core. However, for the very heavy NSs, heaviest so far observed having mass around $M \sim 2 M_{\astrosun}$ \cite{Antoniadis:2013pzd, Demorest:2010bx} the existence of an exotic inner core is assumed where additional degrees of freedom can appear, such as hyperons or quark matter. So far there is no direct evidence for the inner crust composition of NSs so the questions remains open for further observational and computational efforts.
The observational efforts to know more about NS's are plentiful, coming from for e.g. NICER made to explore NS composition, or orbiting X-ray satellites such as Chandra that bring a wealth of information on nuclear reactions thought to occur in the extreme, high-density environments of NS's. Moreover, we have already begun to study some of these processes in our laboratories with the existing (NSCL-MSU, GANIL, GSI, RIKEN) and future major facilities (FAIR (GSI), SPIRAL2 (GANIL), HIE-ISOLDE (CERN), FRIB (MSU), ARIEL (TRIUMF), KORIA (South Korea), CARIF (China) etc. (see Ref.\cite{Blumenfeld:2013jza} for an overview). The future radioactive ion beam facilities will provide research opportunities not available with ordinary ion beams, allowing the investigation of nuclear reactions important to the stellar burning and nucleosynthesis which occur in high temperature and/or density environments in stars. They will allow the study of very neutron rich nuclei close to the neutron drip line in the valley of stability. Knowing the precise masses of these nuclei has also an influence on the composition of NS crusts, whose structure and properties are important for many aspects in NSs and supernova physics, such as heat transport and neutrino opacities.

\begin{table*}[t] 
	\centering
	\caption{The values of QMC$\pi$-II model free parameters: coupling constants for different mesons ($G_m$ for $m=\sigma$, $\omega$, $\rho$), $\sigma$-meson mass ($m_{\sigma}$), pairing parameters ($V_p$, $V_n$) and $\sigma$-meson self-interaction parameter ($\lambda_3$), fitted to experimental data from Wang and Audi \cite{Wang_2017}.}
	\label{tab:QMCpi2}       
	\begin{tabular}{l c c c c c c c c c c c }
		\hline
		Parameter& &$G_{\sigma}$ & $G_{\omega}$ & $G_{\rho}$ & &$m_{\sigma}$ & &$V_p$ & $V_n$ & &$\lambda_3$ \\ 
		& & $[fm^2]$ & $[fm^2]$ & $[fm^2]$ & & $[MeV]$ & & $[MeV]$ & $[MeV]$ & &	\\\hline
		Value & &$9.046$ & $5.287$& $4.708$ && $494.7$ && $287.5$ & $275.001$ && $0.049$ \\
		\hline
	\end{tabular} 
\end{table*}

In Section 2, we will introduce the theoretical model used in this work to describe properties of nuclear matter inside of NSs, the Quark-meson coupling (QMC) model. Neutron star structure is discussed in Section 3. The model is used to calculate the outer crust properties while remarks are made on the composition of neutron star inner crust and core. Here we also refer to the previous calculations for the heavy NSs
done within our working group at University of Adelaide and with our collaborators. Finally, the conclusions are drawn and the importance of developing a unique equation of state (EOS) for NSs within the same model is emphasised in Section 4.

\section{Quark-meson coupling model}\label{sec:2}

There are different approaches to describe dense nuclear matter. In ideal case, we would use quantum chromodynamics (QCD) to solve the problem of quarks and gluons using lattice calculations. However, this approach is still limited to few nucleon systems. For heavier nuclei, the ab-initio approaches are used which assume realistic nucleon-nucleon (NN) interaction designed to describe NN scattering in vacuum and properties of light nuclei. Those methods are numerically complex, which limits them to the proximity of magic shell closures or up to a certain mass number. For a universal approach the phenomenological models based on effective interactions are used, depending on a number of parameters (usually $10-15$) fitted to nuclear matter properties and selected properties of several nuclei over the full nuclei chart. These are interpreted in terms of energy density functional (EDF) theory and are most widely used methods in the construction of astrophysical EOSs. 

The QMC model is also in this category but in contrast with other models it takes into account the internal quark structure of a nucleon. In this formulation nucleons are treated as confined, non-overlapping bags of three quarks where the interaction is modelled through the exchange of effective mesons ($\sigma$, $\omega$ and $\rho$) between quarks from different bags, which are, in turn, modelled using the MIT bag model. This approach self-consistently relates nucleon's dynamic to the relativistic mean fields expected to arise in nuclear matter. The motivation to develop this kind of approach came from an effort to explain the European Muon Collaboration (EMC) effect \cite{AUBERT1983275, Geesaman:1995yd}, the surprising observation that the cross section for deep inelastic scattering from an atomic nucleus is different from that of the same number of free nucleons.
 
The basic idea is that the application of scalar mean field $\sigma$ to a bound nucleon can lead to significant changes in nucleon's structure due to the $\sigma$ field strength that can be up to half of a nucleon mass. This effect is parametrized in terms of "scalar polarizability" $d$, introducing an additional term in nucleon effective mass 
\begin{equation}
M^{\ast}_N = M_N - g_{\sigma}\sigma + \frac{d}{2}(g_{\sigma}\sigma)^2,
\end{equation}
where $g_{\sigma}\sigma$ is the strength of the applied scalar field.

The original derivation of an EDF based on the QMC model was given in Ref.\cite{Guichon:2006er} and a comprehensive survey of the theory, the latest developments and applications
of the model, can be found in the recent review paper of 2018 \cite{GUICHON2018262}. The latest version of the model, QMC$\pi$-II \cite{Martinez:2018xep}, is developed in effort to obtain the values od incompressibility K and slope of the symmetry energy L for nuclear matter closer to the range of generally accepted values ($K = 248\pm8$ MeV \cite{Piekarewicz:2003br} or $K=(240\pm20)$ MeV \cite{Shlomo2006}, and $L = (58.7\pm 28.1)$ MeV \cite{RevModPhys.89.015007}), than previous QMC-I-$\pi$ model \cite{Stone2016PRL}. For that reason, QMC$\pi$-II introduces higher order self-interaction of the $\sigma$ meson, expanding the $\sigma$ field potential energy by including the cubic term
\begin{equation}
V(\sigma) = \frac12 m_{\sigma}^2 \sigma^2 + \frac{\lambda_3}{3!}(g_{\sigma} \sigma)^3.\\
\end{equation}
The additional self-interaction parameter $\lambda_3$ is introduced and fitted to the experimental data which brings us to seven free parameters in total: $\lambda_3$, the $\sigma$ meson mass ($m_{\sigma}$), the effective meson coupling constants $G_{m}= g^2_m / m^2_{m}$ (where $m$ stands for different mesons $m = \sigma, \omega, \rho$) and the two pairing strengths $V_p^{pair}$ and $V_n^{pair}$. These are fitted to 163 data points which include binding energies, root mean-square charge radii and proton and neutron pairing gaps of seventy magic and doubly magic nuclei. The fitting procedure of model parameters is described in detail in Ref.\cite{Martinez:2018xep}. The final set of parameter values is given in Table \ref{tab:QMCpi2}. 

The introduction of higher order self-interaction of the $\sigma$ meson does improve the calculated K and L values compared to previous QMC-I-$\pi$ model and brings them in the range of experimentally expected values, as given in Table \ref{tab:KandL}.

\begin{table}[b] 
	\centering
	\caption{The values of incompressibility K and slope of symmetry energy L for previous and latest QMC models compared to their generally accepted values coming from various experiments.}
	\label{tab:KandL}       
	\begin{tabular}{l c c c}
		\hline
		Parameter& Exp & QMC-I-$\pi$ & QMC$\pi$-II  \\ \hline
		$K$ $[MeV]$ & $240\pm20$  \cite{Shlomo2006} & $319$ & $270$	\\
		$L$ $[MeV]$ & $58.7\pm 28.1$  \cite{RevModPhys.89.015007}  & $17$ & $70$	\\\hline
		\hline
	\end{tabular} 
\end{table}

\begin{figure*}[!tb]
	\centering
	\includegraphics[scale=0.6]{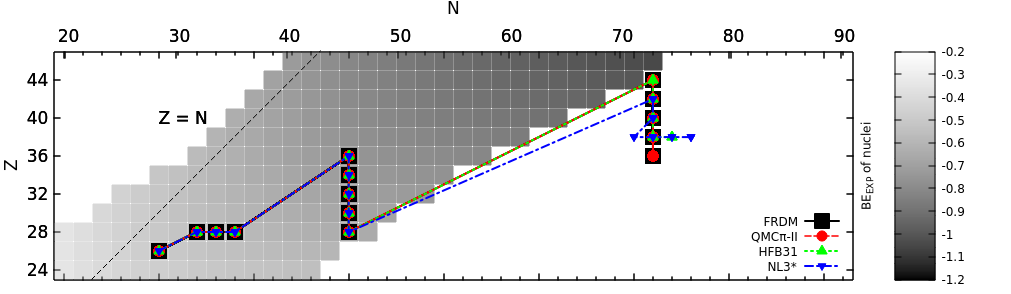}
	\caption{The sequence of nuclei appearing in the NS's outer crust given in Z-N space. The QMC$\pi$-II model predictions are compared to several other models: FRDM\cite{MOLLER1995185,MOLLER1997131}, HFB31 \cite{PhysRevC.93.034337, BRUSLIB} and NL3* \cite{PhysRevC.89.054320}.}
	\label{fig:EOS}
\end{figure*}

\section{Neutron star structure} \label{sec:3}

The structure of a static, non-rotating NS at $T=0$ is presented and the properties of each layer are discussed in the following.

\subsection{Outer crust}
In the following we are adopting the approach of Ref.\cite{PhysRevC.83.065810} in order to model the outer crust of a NS. Starting with a lattice of nuclei at its surface, immersed in the free electron gas, we determine the nuclei species appearing throughout the crust by minimizing the Gibbs free energy per nucleon, $g$, defined as
\begin{equation}\label{eq:Gibbs}
g = e + \frac{P}{n},
\end{equation}
where $e$ is the energy per nucleon, $P$ is the total pressure at the given point in NS crust and $n$ is the nuclear matter density calculated numerically for a given pressure. The energy per nucleon in the outer crust region is
\begin{equation}\label{eq:Mprime}
e = M'(A,Z)/A + e_e + e_L
\end{equation}
where $M'(A,Z)$ is the atomic mass of the nucleus where the binding energy of the atomic electrons is subtracted out , $e_e$ is the electron contribution and $e_L$ comes from the lattice. The only necessary input to the calculation are the nuclear masses. However, the very neutron rich nuclear masses are not experimentally available. These are calculated through theoretical models, in this case the QMC$\pi$-II model.

The assumption is that only one species of nucleus exists for each value of pressure, resulting in density discontinues between two different nuclei that appear in sequence with increasing density. These intervals of two coexisting nuclei are excluded from this approach.

The sequence of outer crust nuclei is given in Fig.\ref{fig:EOS} where QMC$\pi$-II prediction is compared to the calculations of finite range droplet model (FRDM) \cite{MOLLER1995185,MOLLER1997131} for which the latest FRDM(2012) parametrisation is used  \cite{MOLLER20161}, to the Skyrme type model HFB-31 \cite{PhysRevC.93.034337, BRUSLIB} and the relativistic mean-field type, the NL3* \cite{PhysRevC.89.054320}. Up to $^{78}Ni$ experimental values for nuclear masses are used. For following nuclei experimental data is not available and therefore theoretical models are used to deduce the nuclei masses, resulting in slight differences in the nuclei appearing in the sequence. The QMC$\pi$-II model predicts the following sequence: $^{56}Fe$, $^{62}Ni$, $^{64}Ni$, $^{66}Ni$, $^{84}Kr$, $^{82}Ge$, $^{80}Zn$ and $^{78}Ni$, $^{126}Ru$, $^{124}Mo$, $^{122}Zr$, $^{120}Sr$ and $^{118}Kr$. The uncertainty of the nuclei building the bottom of the outer crust can be avoided by the mass measurements of these very neutron rich nuclei, which should be possible with the next generation facilities listed in Section \ref{sec:1}.

\subsection{Inner crust}\label{subsec:3.1}

Modelling of the inner crust of NSs is far more complicated. We do not simply have different nuclear species in lattice arrangement with free electron gas as a background but also free neutron gas where neutrons interacting with nuclei and each other. This makes the inner crust of a NS an unique system, not accessible in the laboratories, which means we rely completely on theoretical models. As the density increases it is assumed that nuclei clump into spherical forms and with further increase of density various shapes can appear (rods, slabs, tubes etc.), commonly called nuclear pasta. Modelling of pasta phases is a demanding task but also very important since this region has a significant influence on the NS EOS. A lengthy overview of the models used in this region and their predictions is given in Ref.\cite{Chamel:2008ca}. Within the QMC model, the inner crust is yet to be modelled.

\subsection{Neutron star core} \label{subsec:3.2.}
 
 The NSs were modelled previously with QMC model \cite{RIKOVSKASTONE2007341}, producing the two solar mass  NS even before the first one was observed and also while including hyperons. The latest QMC calculation additionally includes the nucleon-nucleon interaction via the exchange of $\delta$ meson, which can have an influence on the predicted NS radius \cite{Motta:2019tjc}. The $\delta$ meson was previously often ignored in NS calculations where the main constraint was to reach the $2 M_{\astrosun}$ mass, since it doesn't influence the NS maximum mass prediction. However, with the gravitational wave observation of GW170817, the radius also becomes relevant and the influence of the $\delta$ meson on its prediction with the QMC$\pi$-II model was investigated \cite{Motta:2019tjc}. 

\section{Conclusion}\label{sssec-2.1.2}

The QMC model that takes into account the nucleon structure is a promising approach to describe not only nuclear matter but finite nuclei. It can account for the EMC effect [24] as well as predict new observables (spin EMC and isovector EMC effects) [25, 26] through which it can be tested. Within this work we have shown that by taking quarks as degrees of freedom and having different saturation mechanism we can obtain the same quality of results as the other commonly used nuclear matter models while having less parameters. The necessity for mass measurements of heavier neutron rich nuclei, in order to predict the outer crust composition with certainty, is demonstrated. 

The goal would be to construct the EOS of nuclear matter that could describe the full density range found in NS, from very dense NS cores to the outer crust. With the present work, another step in that direction is made. The model is currently going through further extensions in several directions, first to improve the prediction of nuclei's binding energies and charge radii with the inclusion of tensor terms in the QMC EDF (the QMC$\pi$-0 model) but also in direction of making the model temperature dependent. The temperature dependence of QMC model will enable application to proto-neutron stars \cite{Stone:2019abq}. The effort in both directions is in progress.

\section{Acknowledgements}

The author S.A. thanks the organisers of HIAS2019 for the symposium invitation and to Dr. Cedric Simenel attendance support through the Australian Research Council discovery projects grant DP18.

\bibliographystyle{unsrt}
\bibliography{SAntic-hias2019}

\end{document}